\documentclass[12pt]{article}
\usepackage[pctex32]{graphics}
\begin{document}
~
~
\vspace{1cm}
\begin{center} {\Large \bf  Fuzzy BIons in Curved Spacetimes}
                                                  
\vspace{1cm}

                      Wung-Hong Huang\\
                       Department of Physics\\
                       National Cheng Kung University\\
                       Tainan, Taiwan\\

\end{center}
\vspace{1cm}
\begin{center} {\large \bf  Abstract} \end{center}
The non-abelian Dirac-Born-Infeld action is used to construct the D2-brane from multiple D0-branes in the curved spacetimes.  After choosing the matrix elements as the coordinates of the D0-branes we obtain a simple formula of the Lagrangian for the system in a class of the curved background.  Using the formula we first re-examine the system in the flat spacetime and show that, in addition to the fuzzy tube and fuzzy spike which were found in the previous literature, there is the fuzzy wormhole solution.  Next, we apply the formula to the system in the geometry of the NS5-branes background.  A solution describing  the fuzzy BIon of spike profile is obtained.  Our investigations show that the size of the matrices is finite for the fuzzy spike in the curved spacetimes.

\vspace{2cm}
\begin{flushleft}
E-mail:  whhwung@mail.ncku.edu.tw\\
\end{flushleft}
\newpage
\section {Introduction}
Myers effect [1] shows that, in the background of RR potential  the multiple D0-branes will expand into a fuzzy D2-brane with a spherical profile.   This is one of the general properties that the D(p+2)-brane could be realized by the fuzzy configurations of the Dp-branes [2-4].  The phenomena are described by the   non-abelian Dirac-Born-Infeld action [1] and have been extended to the system in the curved spacetimes [3], in which the fuzzy configurations of sphere and cylinder have been studied.

  The fuzzy spike in the curved background, in which the ``radius" $r(z)$  depends on the coordinate of $z$ direction has not yet been discussed.  The problem of finding the solution is investigated in this paper.  We will see that, although a simple formula of the Lagrangian for the multiple D0 in a class of the curved background could be obtained it is difficult to obtain the exact solution therein.  A solution we found is the system in the geometry of the NS5-branes background.   From our investigations it indicates that the size of matrices for a fuzzy spike is finite when it is in the curved spacetimes.  We also in this paper re-examine the system in the flat spacetime and show that, in addition to the fuzzy tube and fuzzy spike which were found in the previous literature, there is the fuzzy wormhole solution.   Note that using the abelian Dirac-Born-Infeld action the BIons with spike and wormhole profiles have been found by Emparan [5].  Our investigations in his paper have used the method of Hyakutake [2] while extend it to the curved spacetimes.

    In section II we simplify the non-abelian Dirac-Born-Infeld action in a class of the curved background.  The detailed calculations are present.  After choosing the matrix elements as the coordinates of the D0-branes we obtain a simple formula of the Lagrangian for the system in the class of the curved background.  Using the formula we first in section III review the fuzzy solutions in flat spacetime, i.e.  the fuzzy tube and fuzzy spike which were found in the previous literature.  We then present a new fuzzy wormhole solution.   In section IV, we apply the formula to the system in the geometry of the NS5-branes background.  We present a solution of fuzzy BIon of spike profile.  From our investigations it is seen that the size of the matrices is finite for the fuzzy spike in the curved spacetimes. We make a discussion in the last section.  Note that in recent many authors have investigated the dynamics of D-branes in the NS5-branes [6,7].  Their results are used to show that the radial mode of the BPS D-brane in the NS5-branes backgrounds resembles the tachyon rolling dynamics of unstable D-brane [8]. The solution found in this paper could be used to discuss the relevant problem.

\section {Non-abelian Dirac-Born-Infeld Action in Curved Spacetimes}
The Myers T-dual non-abelian Born-Infeld action describing N coincident Dp-branes is given by [1].   
$$S_{BI} = -T_p \int d^{p+1}\sigma\, STr\left(e^{-\phi}\sqrt{-det\left(P\left[E_{ab}+E_{ai}(Q^{-1}-\delta)^{ij}E_{jb}\right] + \lambda F_{ab}\right)\, det(Q^i{}_j)} \right) ,   \eqno{(2.1)}$$
in which $F_{ab}$ is the DBI 2-form strength on the Dp-brane and 
$$ E_{\mu\nu} = G_{\mu\nu}+B^{NS}_{\mu\nu}, \hspace{1.7cm}   \eqno{(2.2)}$$
$$ Q^i{}_j = \delta^i{}_j+{i\over \lambda}~ [X^i, X^k]~ E_{kj},  \eqno{(2.3)}$$
where $G_{\mu\nu}$ is the metric of the spacetime, $B^{NS}_{\mu\nu}$ the NS 2-form, and $X^i$ are the $N \times N$ matrices describing the coordinates of the N Dp branes. The pull-back of the bulk spacetime tensors $V_{\mu_1 \cdot\cdot\cdot\mu_n}$ to the D-brane world-volume is denoted by the symbol $P[V_{\alpha_1...\alpha_n}]$ in which   
$$ P[V_{\alpha_1...\alpha_n}]= V_{\mu_1 ..\mu_n} D_{\alpha_1} X^{\mu_1} \cdot\cdot\cdot D_{\alpha_n} X^{\mu_n},  \eqno{(2.4)}$$
with $D_a X^i = \partial_aX^i + i[A_a,X^i]$ .  The brane tension is defined by
$$ T_p={2\pi\over g_s\left(2\pi\ell_s\right)^{p+1}}, \eqno{(2.5)}$$
and  $\lambda \equiv 2\pi \ell_s^2$ in which $\ell_s$ is the string length scale.  

  A class of curved spacetime we will consider is described by 
$$G_{\mu\mu} = \left(g_{00},\,g_{xx},\,g_{yy},\,g_{zz}\right), \eqno{(2.6a)}$$
$$ {\bf B}^{NS} = B_{0z}\,dt \wedge dz. \hspace{1.5cm} \eqno{(2.6b)}$$
In this paper we will consider the system without the RR potential and thus the Wess-Zumino term [1] does not appear. 

  For the case of N coincident D0-branes in the class of the curved background described by (2.6) the relevant quantities become

$$Q^i{}_j = \left( \begin{array}{ccc}
1&  i \lambda^{-1} [X^1, X^2] \, g_{yy}& i \lambda^{-1} [X^1, X^3] \, g_{zz}\\
 i \lambda^{-1} [X^2, X^1] \, g_{xx}&1& i \lambda^{-1} [X^2, X^3] \, g_{zz}\\
 i \lambda^{-1} [X^3, X^1] \, g_{xx}&i \lambda^{-1} [X^3, X^2] \, g_{yy}&1
\end{array}
\right). \eqno{(2.7)}$$
\\
$$det(Q^i{}_j) \stackrel{STr}{=} 1- \lambda^{-2} [X^1, X^2]^2 g_{xx}g_{yy} -\lambda^{-2} [X^2, X^3]^2 g_{yy}g_{zz}- \lambda^{-2} [X^3, X^1]^2 g_{zz}g_{xx}, \eqno{(2.8)}$$
The notation $\stackrel{STr}{=}$ is used to emphasize that the above equation holds under the symmetrized trace prescription [1].  The pull-back of the bulk spacetime tensors are
$$P[E_{00}]= g_{00} - g_{xx} [A_0, X^1]^2 - g_{yy}[A_0, X^2]^2 - [A_0, X^3]^2 g_{zz}. \hspace{3.7cm} \eqno{(2.9)}$$
$$P\left[E_{0i}(Q^{-1}-\delta)^{ij}E_{j0}\right] = - B_{0z}(Q^{-1}-\delta)^z{}_z\, g^{-1}_{zz} + i B_{0z} (Q^{-1}-\delta)^z{}_i\, [A_0, X^i]  \hspace{2cm}$$
$$ \hspace{4cm}-  i B_{0z} E_{ki}(Q^{-1}-\delta)^i{}_z\, [A_0, X^k] \, g^{-1}_{zz} - E_{mi}(Q^{-1}-\delta)^i{}_n\, [A_0, X^m] \, [A_0, X^n], \eqno{(2.10)}$$
and finally we have a simple relation
$$ det\left(P\left[E_{00}+E_{0i}(Q^{-1}-\delta)^{ij}E_{j0}\right]\right)det(Q^i{}_j)  = g_{00} - g_{00}g_{xx} g_{yy}\,\lambda^{-2}[X^1,X^2]^2\hspace{3cm} $$
$$\hspace{2cm}  - g_{00} g_{yy} g_{zz}\,\lambda^{-2} [X^2,X^3]^2  - g_{00} g_{zz} g_{xx} \,\lambda^{-2}[X^3,X^1]^2  -  g_{yy}B_{0z}^2 \,\lambda^{-2}[X^2,X^3]^2 $$
$$  \hspace{2.8cm} -  g_{xx}\,\lambda^{-2} B_{0z}^2 [X^3,X^1]^2 - g_{xx} [A_0,X^1]^2 - g_{yy} [A_0,X^2]^2 - g_{zz} [A_0,X^3]^2. \eqno{(2.11)}$$
in which we have neglected terms containing higher power of commutative matrices.   This is the first generally useful formula presented in this paper.  Note that many interesting backgrounds fall into the class of eq.(2.6), for example, the $AdS_3 \times S^3$, NS5-branes and macroscopic string backgrounds. 

 To proceed, we will consider the fuzzy surface with axial symmetry around the $x^3 (= z)$ direction, thus the matrices is chosen as [2]
$$  X^1_{mn} = \frac{1}{2} \rho_{m+1/2} \delta_{m+1,n}  + \frac{1}{2} \rho_{m-1/2} \delta_{m,n+1},\eqno{(2.12a)}$$
$$ X^2_{mn} = \frac{i}{2} \rho_{m+1/2} \delta_{m+1,n} 
  - \frac{i}{2} \rho_{m-1/2} \delta_{m,n+1} ,  \eqno{(2.12b)}$$
$$  X^3_{mn} = z_m \delta_{m,n} , \hspace{4.5cm}\eqno{(2.12c)}$$
where $m,n$ are the set of integers.   $z_m$ is interpreted as a position of the $m$th segment in the $z$ direction.  Because that 
$$\left(X^1\right)^2_{mn} + \left(X^2\right)^2_{mn}= {1\over2}\left(\rho_{m+1/2}^2 +\rho_{m-1/2}^2 \right)\delta_{m,n} \equiv r_m^2 \, \delta_{m,n},\eqno{(2.13)}$$
the function  $r_m\equiv {1\over 2}\,\left(\rho_{m+1/2}^2+\rho_{m-1/2}^2\right)^{1/2}$\, is naturally interpreted as a position at  $m$th segments in the radial direction.   Values of $m,n$ run an infinite set of integers for the open surface and a finite
set of integers for the closed surface. A possible fuzzy spike is plotted in figure 1. 
\\
\\
\scalebox{1}{\hspace{6cm}\includegraphics{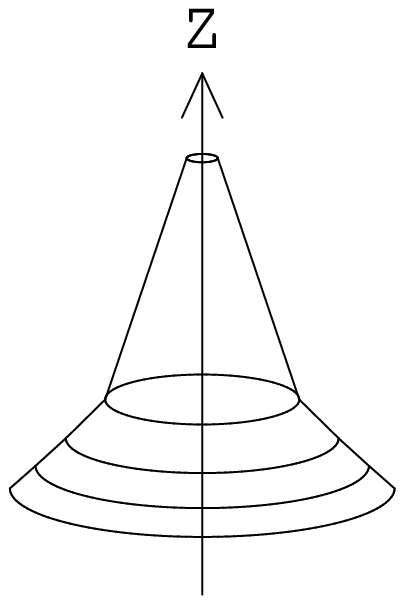}}
\\
{. \hspace{4cm} Figure 1. The profile of a fuzzy spike.}
\\
\\
 The commutation relations of  $X^i$ are evaluated as
$$ [X^1,X^2]_{mn} = - \frac{i}{2} \big( \rho_{m+1/2}^2 \!-\! 
  \rho_{m-1/2}^2 \big) \delta_{m,n}, \eqno{(2.14a)}$$
$$[X^2,X^3]_{mn} = \frac{i}{2} \rho_{m+1/2} \big( z_{m+1} \!-\! z_m \big) 
  \delta_{m+1,n}  + \frac{i}{2} \rho_{m-1/2} \big( z_{m} \!-\! z_{m-1} \big) \delta_{m,n+1} , \eqno{(2.14b)}$$
$$ [X^3,X^1]_{mn} = - \frac{1}{2} \rho_{m+1/2} \big( z_{m+1} \!-\! z_m \big)   \delta_{m+1,n}  + \frac{1}{2} \rho_{m-1/2} \big( z_{m} \!-\! z_{m-1} \big) \delta_{m,n+1}. \eqno{(2.14c)}$$
Thus 
$$ \left([X^1,X^2]\right)^2_{mn} = - \frac{1}{4} \big( \rho_{m+1/2}^2 \!-\! 
  \rho_{m-1/2}^2 \big)^2 \delta_{m,n}, \hspace{6cm}\eqno{(2.15a)}$$
$$\left([X^2,X^3]\right)^2_{mn} = \left([X^3,X^1]\right)^2_{mn}= - \frac{1}{4}\left[\rho_{m+3/2} \rho_{m+1/2}(z_{m+2}-z_{m+1})(z_{m+1}-z_m) \delta_{m+2,n} \right.$$
$$\hspace{6cm}+ \left(\rho_{m+1/2}^2(z_{m+1}-z_m)^2 +  \rho_{m-1/2}^2(z_{m}-z_{m-1})^2\right) \delta_{m,n} $$
$$\left.\hspace{7cm}+\rho_{m-1/2} \rho_{m-3/2}(z_{m}-z_{m-1})(z_{m-1}-z_{m-2}) \delta_{m-2,n}\right] , \eqno{(2.15b)}$$
We will consider the system  with  
$$A_0 = a_0 X^3. \eqno{(2.16)}$$
This means that there is an electric flux along the $z$ direction and the system have the fundamental strings along the z axial. 

 Substituting (2.15) into (2.11)  the Lagrangian for the N coincident D0-branes in the  class of the curved background becomes
$$ L\! = \!T_0 \, e^{-\phi}\left(\!-N\left(\!-\!g_{00}\right)^{1/2}\! - \!\sum_m\left[ {\lambda^{-2}\over 4}\left(\!-\!g_{00}\right)^{1/2}\left({1\over 2} g_{xx} g_{yy}\big( \rho_{m\!+\!1/2}^2 \!-\!\rho_{m\!-\!1/2}^2 \big)^2 \! +\! (g_{yy} g_{zz}\!+\! g_{zz} g_{xx}) \right.\right.\right.$$
$$\times\left. \left. \left.\rho_{m+1/2}^2 \big( z_{m+1} - z_m \big)^2 \right) - {1\over 4}\left(-g_{00}\right)^{-1/2} \left(\lambda^{-2}B_{0z}^2 + a_0^2 \right) \left(g_{xx}+ g_{yy}\right)  \rho_{m+1/2}^2 \big( z_{m+1} \!-\! z_m \big)^2\right]\right).\eqno{(2.17)}$$
This relation is the main formula using in this paper.

  Notice that in the curved spacetime the metric $g_{\mu\nu}$ in (2.17) will be the function of the spacetime.  For the fuzzy surface with axial symmetry around the $z$ direction the metric will depend on the radius $r$ which is defined by  
$$ r^2 = {1\over N}\,Tr \left(\left(X^1\right)^2 + \left(X^2\right)^2\right) = {1\over N}\,\sum_m\left[{1\over 2}\left(\rho_{m+1/2}^2+\rho_{m-1/2}^2\right)\right] \equiv {1\over N}\,\sum_m \, r_m^2\,,\eqno{(2.18)}$$
in which we have used the relation (2.13) in which $r_m$\, is interpreted as the  position at $m$th segment in the radial direction.
\section {Fuzzy BIons in Vacuum}
In the vacuum the main formula (2.17) becomes
$$ L =  \sum_m\left[ - {\lambda^{-2}\over 8} \big( \rho_{m+1/2}^2 -  \rho_{m-1/2}^2 \big)^2  - {\lambda^{-2}\over 2} \rho_{m+1/2}^2 \big( z_{m+1}- z_{m}\big)^2 + {1\over 2}a_0^2  \rho_{m+1/2}^2 \big( z_{m+1}- z_{m}\big)^2 \right].\eqno{(3.1)}$$
After the variation with respective to the $\rho_{m+1/2}$ and $z_m$ we have the equations
$$ \big( \rho_{m+3/2}^2 - 2 \rho_{m+1/2}^2 + \rho_{m+1/2}^2 \big) - 2 (1 - a_0^2\lambda^{2}) \big( z_{m+1}- z_{m+1}\big)^2  = 0. \eqno{(3.2)}$$
$$ \rho_{m+1/2}^2 \big( z_{m+1}- z_{m}\big) - \rho_{m-1/2}^2 \big( z_{m1}- z_{m-1}\big) = 0. \eqno{(3.3)}$$
Searching the solution from above equations is an interesting problem and the result represents  the fuzzy BIon in the vacuum.   Note that the above equations had been obtain by Hyakutake [2] from the SFSS matrix theory. 

  There are two nontrivial solutions presented in the previous literature.

 (I) Fuzzy tube : The fuzzy tube solution is described by  
$$\rho_{m+1/2} = r, ~~~~~ z_m = L m. \eqno{(3.4)}$$
In this case, the fuzzy tube has a constant radius while its position of the $m$th segment in the $z$ direction is a discrete value with equal spacing.  The solution of eq.(3.4) implies that the commutation relations (2.14) become
$$[X^1,X^2] = 0, ~~ [X^2,X^3] = iL X^1,~~~~[X^3,X^1] = iL X^2, \eqno{(3.5)}$$
which is just the matrix tube found in [9].    The tube solution was first found in [10] by using the abelian Dirac-Born-Infeld action.  Note that Eq.(3.4) shows that $ -\infty < m < \infty  $, thus the size of the matrices is infinite.

 (II) Fuzzy spike : The fuzzy  spike found in [2] is a solution of  the following matrix elements
$$\rho_{m+1/2} = 2\alpha m,\eqno{(3.6a)}$$
$$ \hspace{2cm} z_m = c + L \sum_{i=1}^{m-1}{1\over i},\eqno{(3.6b)}$$
$$~~~~~~ a_0 = \lambda^{-1}.\eqno{(3.6c)}$$
Thus, increasing the $m$ the  position $z_m$ is increasing and the radius of the BIon increasing also.  The  profile of BIon is spike as that plotted in figure 1.  Eq.(3.6b) shows that $ 1 < m < \infty  $, thus the size of the matrices  of fuzzy spike is infinite.

 (II) Fuzzy wormhole : We now present the fuzzy wormhole solution.  It is described by the following matrix elements
$$\hspace{2cm}\rho_{m+1/2} = (1-\lambda^2\,a_0^2)L^2 \left(m+{3\over2}\right)^3\left(m+ {7\over2}\right),\eqno{(3.7a)}$$
$$ z_m = c + L \sum_{i=m_0}^{m}\left( i+{1\over2}\right),\eqno{(3.7b)}$$
in which $m_0$ is an arbitrary integral. Note that form (3.7a) we find that the radius ``$r_m$"  at $m$th segment (defined in (2.13)) is  
$$r_m^2 = {1\over 2} L^2 \left[ (m+{5\over2})^3(m+{7\over2})+(m+{3\over2})^2(m+{5\over2}) \right]\eqno{(3.8)}$$
It is an easy work to check that $r_m^2 > 0 $ for any integral $m$, thus the size of the matrices  of fuzzy wormhole is infinite.  Eq.(3.8) show that $r_m \rightarrow \infty$ at $m = \pm \infty$ and a minimum is at $m = 0$, thus the profile of the fuzzy BIon is a wormhole.

\section {Fuzzy BIons in NS5-branes Background}
In this section we will use the main formula (2.17) to find the fuzzy spike in NS5-branes background.  The background fields around $N_5$ NS5-branes are given by the CHS solution [11]. The metric, dilaton and NS-NS $B^{NS}$ field are
$$ds^2 = dx_\mu dx^\mu + h(x^n) dx^mdx^m ,$$
$$e^{2(\phi)}=h(x^n)\, ,~~~~~~~~~~~~~~~~~$$
$$H_{mnp} = -\epsilon_{mnp}^q\partial_q\phi\,.~~~~~~~~~~~~~~~~ \eqno{(4.1)}$$
Here $h(x^n)$ is the harmonic function describing fivebranes, and $H_{mnp}$ is the field strength of the NS-NS $B^{NS}$ field. For the case of coincident $N_5$ fivebranes one has
$$h(r) =1 + {N_5 \over r^2} , \eqno{(4.2)}$$
where $r = |\vec x|$ is the radial coordinate away from the fivebranes in the transverse space labeled by $(x^6,\cdots, x^9)$.  

   Using the above metric and fields we can easily from (2.17) show that the Lagrangian  of $N$ coincident D0-branes in the NS5-branes background becomes 
$$ L =  - {N\over \sqrt h} -{\lambda^{-2}\over 8}\, h\, \sqrt h \,\sum_m\left(\rho_{m+1/2}- \rho_{m-1/2}\right)^2  + {1\over 2}\left(a_0^2-\lambda^{-2} \right) \, \sqrt h \,\sum_m \rho_{m+1/2}^2 \big( z_{m+1}- z_{m}\big)^2. \eqno{(4.3)}$$
After the variation with respective to the $\rho_{m+1/2}$ and $z_m$ we have the equations
$$ -{\partial\over r \partial r}\left({N\over \sqrt h}\right) - {\partial\over r \partial r}\left(h\, \sqrt h\right) {\lambda^{-2}\over 8}\,\sum_m\left(\rho_{m+1/2}- \rho_{m-1/2}\right)^2 - {\lambda^{-2} \over 2} h\, \sqrt h\,\big( \rho_{m+3/2}^2 - 2 \rho_{m+1/2}^2 + \rho_{m+1/2}^2 \big)$$
$$+ {1\over 2}\left(a_0^2-\lambda^{-2} \right) \left[{\partial\over r \partial r}\left(\sqrt h\right)\left(\sum_m \rho_{m+1/2}^2\big( z_{m+1}- z_{m+1}\big)\right)  + 2 \sqrt h \big( z_{m+1}- z_{m+1}\big)^2\right] = 0. \eqno{(4.4)}$$
$$ \rho_{m+1/2}^2 \big( z_{m+1}- z_{m}\big) - \rho_{m-1/2}^2 \big( z_{m1}- z_{m-1}\big) = 0. \eqno{(4.5)}$$
The fuzzy  spike in the NS5-branes background can be obtained by choosing the following matrix elements
$$\rho_{m+1/2}^2 = 2 \alpha (m +\beta),\eqno{(4.6a)}$$
$$  z_m = c + L \sum_{i=1}^{m-1}{1\over i+\beta},\eqno{(4.6b)}$$
$$~~~~~~ a_0 = \lambda^{-1}.\eqno{(4.6c)}$$
which becomes the fuzzy  spike solution in the vacuum if $\beta =0$.  Above solution automatically satisfies eq.(4.5) and after substituting them into eq.(4.4) we find that 
$$ r^2 =  {N_5\over \sqrt{8N\lambda^2\over 3\alpha }-1}.\eqno{(4.7)}$$
Now, as shown in (2.18) the value ``r" shall be defined by
$$r^2 = {1\over N}\sum_m \rho_{m+1/2}^2 = {1\over N}\sum_m \alpha (m +\beta),\eqno{(4.8)}$$
the value of  $\beta$ could therefore be found for a fixed $\alpha$.  The value $N$ specifies the size of the matrices of  the fuzzy spike in the curved spacetimes. 

   We now make following comments to discuss the physical properties of the parameters $\alpha$, $L$, and $N$.

(I) Substituting (3.6a) or (4.6a) into (2.14a) we see that 
$$ [X^1,X^2]_{mn} = - i\, \alpha \,\delta_{m,n}. \eqno{(4.9)}$$
Thus, the value $\alpha$ represents the fuzziness of the D0-branes. Comparing to the approach of abelian Dirac-Born-Infeld,  it is known that the value of $\alpha$ shall be proportional to the density of D0-branes, thus it will be proportional to the magnetic flux of the system.

(II) The value of $L$ in (3.6b) or (4.6b) could not be fixed by the equation.   This means that the scale of fuzzy BIon along ``z" axial is arbitrary.  The fact may be traced to the conformal symmetry of the string.  Another arbitrary constant $c$ appearing (3.6b) or (4.6b)  reveals the translation symmetry along the $z$ direction. 

(III) From eq.(4.8) we see that the value $N$, which specifies the size of the matrices of  the fuzzy  spike in the curved spacetimes, shall be a finite number.   This is the general property for a fuzzy spike in the curved background, in contrast to that in flat spacetime (see section III).  This is because that if a fuzzy spike in the curved background has infinite dimension, then the value $\rho_{m+1/2}^2$ shall run to infinite and thus the definition of  ``r" in eq.(4.8) will become infinite in general.   However, the value of  ``r" is a variable  of metric $g_{\mu\nu}$ and it shall be a finite value, thus we conclude that the size of the matrices shall be finite for the fuzzy spike in the curved spacetimes.  It is easy to see that the some conclusion could also be found for the fuzzy wormhole in the curved spacetimes, if it exist.

\section{Conclusion}
In this paper we use the non-abelian Dirac-Born-Infeld action to find the fuzzy  spike solution in NS5-branes background.  We have simplified the non-abelian Dirac-Born-Infeld action in a class of the curved background and presented the detailed calculations to obtain the simple form of the Lagrangian.  Using the formula, after reviewing the solutions of fuzzy tube and fuzzy spike in flat spacetime, which were found in the previous literature,  we first present a new fuzzy wormhole solution.   We next apply the formula to the system in the geometry of the NS5-branes background and present a fuzzy BIon of spike profile.  Our investigations indicates that the size of the matrices is finite for the fuzzy spike in the curved spacetimes.  

 We make following comments to conclude this paper.  

(1) Although the final form of the Lagrangian is obtained in this paper, it is difficult to find the fuzzy BIon solution in general.  For example, we have investigates the system in $AdS_3 \times S^2$ [12] and macroscopic string background [13],  however, the consistent fuzzy spike or wormhole solution has not yet been found.   In fact, even in the NS5-branes background it is difficult to find the fuzzy wormhole solution.   

(2) As mentioned in section I, many authors have investigated the dynamics of D-branes in the NS5-branes, as the results could be used to show that the radial mode of the BPS D-brane in the NS5-branes backgrounds resembles the tachyon rolling dynamics of unstable D-brane [8].  The problem of unstable fuzzy BIons in curved spacetimes have not yet discussed.  It would be interesting to use the solution found in this paper to study the dynamics of fuzzy  spike in the NS5-branes.  The works remain to be studied in the future.

\newpage
{\Large \bf  References}
\begin{enumerate}
\item R.C. Myers, ``Dielectric-Branes'',  JHEP 9912 (1999) 022 [hep-th/9910053];
``Nonabelian Phenomena on D-branes'', Class.Quant.Grav. 20 (2003) S347-S372 [hep-th/0303072]. 
\item Y. Hyakutake, ``Notes on the Construction of the D2-brane from Multiple D0-branes'',  Nucl. Phys.  B675 (2003) 241 [hep-th/0302190]; ``Fuzzy BIon'', Phys.Rev. D68 (2003) 046003 [hep-th/0305019].
\item M. Asano ``Non-commutative branes in D-brane backgrounds'', Int.J.Mod.Phys. A17 (2002) 4733-4748 [hep-th/0106253];  D. K. Park, S. Tamaryan, H. J. W. M\"uller-Kirsten, ``General Criterion for the existence of Supertube and BIon in Curved Target Space'', Phys.Lett. B563 (2003) 224-230 [hep-th/0302145]; Y. Hyakutake, ``Gravitational Dielectric Effect and Myers Effect'' , Phys.Rev. D71 (2005) 046007 [hep-th/0401026];  Wung-Hong Huang ;  ``Fuzzy Rings in D6-Branes and Magnetic Field Background'' , JHEP 0407 (2004) 012 [hep-th/0404202].
\item Y. Kimura, ``Nonabelian gauge field and dual description of fuzzy sphere'',  JHEP 0404 (2004) 058 [hep-th/0402044]; ``Myers Effect and Tachyon Condensation'', Nucl.Phys. B692 (2004) 394-416 [hep-th/0309082]; Y. Hyakutake, ``Torus-like Dielectric D2-brane'',  JHEP 0105 (2001) 013 [ hep-th/0103146]; J. de Boer, E. Gimon, K. Schalm, J. Wijnhout, ``Evidence for a gravitational Myers effect'', Annals Phys. 313 (2004) 402-424 [hep-th/0212250];  B. Janssen, Y. Lozano, ``Non-Abelian Giant Gravitons'', Class. Quant. Grav. 20 (2003) S517-S524 [hep-th/0212257].
\item R. Emparan, ``Born-Infeld strings tuuneling to D-branes'', Phys.Lett. B423 (1998) 71-78 [hep-th/9711106]; C. G. Callen and J.~M.~Maldacena, ``Brane dynamics from the Born-Infeld action'',  Nucl. Phys.  B513 (1998) 198 [hep-th/9708147].
\item D.~Kutasov, ``D-brane dynamics  near NS5-brane,'' [hep-th/0405058]; ``A geometric interpretation of the open string tachyon,'' [hep-th/0408073]; Y.~Nakayama, Y.~Sugawara and H.~Takayanagi, ``Boundary states for the rolling D-brane in NS5 background,'' JHEP  0407(2004) 020 [hep-th/0406173];
D.~A.~Sahakyan, ``Comments on D-brane dynamics near NS5-brane,''
JHEP 0410 (2004) 008 (2004) [hep-th/0408070].
\item  J.~Kluson, ``Non-BPS D-brane near NS5-brane,'' JHEP  0411 (2004) 013 [hep-th/0409298]; ``Non-BPS Dp-brane in the background of NS5-brane on transverse R**3 x S**1,'' [hep-th/0411014]; J.~Kluson and K. Panigrahi, ``Supertube dynamics in diverse backgrounds'', [hep-th/0506012];  S.~Thomas and J.~Ward, ``D-brane dynamics and NS5 rings,'' JHEP 0502 (2005) 015 [hep-th/0411130]; ``D-brane dynamics near compactified NS5-branes'' [hep-th/0501192] ; B.~Chen, M.~Li and B.~Sun, ``D-brane near NS5-brane:  With electromagnetic field,''  JHEP 0412 (2004) 057 [hep-th/0412022];Y.~Nakayama, K.~L.~Panigrahi, S.~J.~Rey and H.~Takayanagi, ``Rolling down the throat  in NS5-brane background: The case of electrified D-brane,'' JHEP 0501 (2005) 052 [hep-th/0412038].
\item A.~Sen, ``Time and tachyon,'' Int.  J.  Mod.  Phys.  A 18 (2003) 4869 [hep-th/0209122];  ``Tachyon matter,'' JHEP  0207  (2002) 065 [hep-th/0203265]; ``Rolling tachyon,'' JHEP  0204 (2002) 048 [hep-th/0203211]; ``Tachyon dynamics in open string theory,'' [hep-th/0410103].
\item D. Bak, K. M. Lee, ``Noncommutative Supersymmetric Tubes'',  Phys. Lett. B509 (2001) 168 [hep-th/0103148]; D. Bak and S. W. Kim, ``Junction of Supersymmetric Tubes,'' Nucl. Phys.  B622 (2002) 95 [hep-th/0108207].
\item D. Mateos and P. K. Townsend, ``Supertubes'', Phys. Rev. Lett. 87 (2001) 011602 [hep-th/0103030]; R. Emparan, D. Mateos and P. K. Townsend, ``Supergravity Supertubes'', JHEP 0107 (2001) 011 [hep-th/0106012].
\item C.~G.~.~Callan, J.~A.~Harvey and A.~Strominger, ``Supersymmetric string solitons,'' [hep-th/9112030].
\item C.~Bachas and M.~Petropoulos, ``Anti-de-Sitter D-branes'', JHEP
{\bf 0102}, 025 (2001) [hep-th/0012234]; P. M. Petropoulos, S. Ribault,  ``Some comments on Anti-de Sitter D-branes'' , JHEP 0107 (2001) 036 [hep-th/0105252];
A.~Giveon, D.~Kutasov and A.~Schwimmer, ``Comments on D-branes in AdS(3)'', Nucl.\ Phys.\ B {\bf 615}, 133 (2001) [hep-th/0106005]; Wung-Hong Huang, ``Anti-de Sitter D-branes in Curved Backgrounds'' JHEP 0507 (2005) 031 [hep-th/0504013].
\item  G.~W.~Gibbons and K.~Maeda, ``Black holes and membranes in higher dimensional theories with dilaton fields,'' Nucl.\ Phys.\ B298 (1988) 741; D.~Bak, S.~J.~Rey and H.~U.~Yee, ``Exactly soluble dynamics of (p,q) string near macroscopic fundamental strings,'' JHEP 0412 (2004) 008 [hep-th/0411099]; Wung-Hong Huang, ``Tubular Solutions in NS5-brane, Dp-brane and Macroscopic String Background'' JHEP 0502 (2005) 061 [hep-th/0502023].

\end{enumerate}
\end{document}